\documentclass[a4paper,11pt]{article}
\pdfoutput=1
\usepackage{jheppub}

\usepackage[utf8]{inputenc}
\usepackage{amsmath}
\usepackage{graphicx}
\usepackage{caption}
\usepackage{subcaption}
\usepackage{amssymb}
\usepackage{color}
\usepackage{cancel}
\usepackage{framed}
\usepackage{comment}

\newcommand{\order}{\mathcal{O}}
\newcommand{\Lagr}{\mathcal{L}}

\newcommand{\nn}{\nonumber}

\title{Collapse of Axion Stars}

\author[1,2]{Joshua Eby,}
\author[1]{Madelyn Leembruggen,}
\author[1]{Peter Suranyi,}
\author[1]{L.C.R. Wijewardhana}
\affiliation[1]{\emph{University of Cincinnati, Dept. of Physics, Cincinnati, OH 45221 USA}}
\affiliation[2]{\emph{Fermi National Accelerator Laboratory, P.O. Box 500, Batavia, IL 60510, USA}}

\emailAdd{ebyja@mail.uc.edu} 
\emailAdd{leembrmn@mail.uc.edu}
\emailAdd{peter.suranyi@uc.edu}
\emailAdd{rohana.wijewardhana@uc.edu}
  
\abstract{Axion stars, gravitationally bound states of low-energy axion particles, have a maximum mass allowed by gravitational stability. Weakly bound states obtaining this maximum mass have sufficiently large radii such that they are dilute, and as a result, they are well described by a leading-order expansion of the axion potential. Heavier states are susceptible to gravitational collapse. Inclusion of higher-order interactions, present in the full potential, can give qualitatively different results in the analysis of collapsing heavy states, as compared to the leading-order expansion. In this work, we find that collapsing axion stars are stabilized by repulsive interactions present in the full potential, providing evidence that such objects do not form black holes. In the last moments of collapse, the binding energy of the axion star grows rapidly, and we provide evidence that a large amount of its energy is lost through rapid emission of relativistic axions.}


\arxivnumber{1608.06911}

\begin{document}
 
\maketitle

\section{Introduction}
The axion, a pseudoscalar particle originally associated with a solution of the strong $CP$ problem in QCD \cite{PQ1,PQ2,Weinberg,Wilczek,DFS,Zhitnitsky,Kim,Shifman}, has been analyzed in a variety of astrophysical contexts, particularly in cosmological evolution \cite{Khlopov1,Khlopov2,Khlopov3,Khlopov4,Khlopov5} and as a candidate for dark matter \cite{Preskill,Sikivie1,Davidson,DF,Holman,Sikivie2}. Axions can condense into gravitationally bound objects, either in the early universe through large-scale overdensities in a coherent axion field (called ``miniclusters''), or through gravitational cooling and collapse (called ``axion stars'') \cite{Tkachev,KolbTkachev,HoganRees}.

The masses of weakly bound axion stars have been computed previously \cite{BB,ESVW}, and they are bounded above by gravitational stability \cite{ESVW,Guth,ChavanisMR}. Axion stars which exceed this maximum mass $M_c$ have a fate which remains an open question. Some authors \cite{Braaten,Braaten2} suggest that such configurations collapse to a compact, very dense state. Recently, the author of \cite{ChavanisCollapse} examined the collapse of a boson star with an attractive self-interaction which has $M>M_c$, using a dynamical equation derived from a Gaussian ansatz for its wavefunction. The author found that, as its potential is unbounded from below, a star of this kind collapses all the way to its Schwarzschild radius and forms a black hole. 

Indeed, the leading axion self-interaction is attractive; axionic or other bosonic objects with repulsive interactions have been considered by \cite{Jiji,EGKW}. However, the axion potential contains additional terms which become increasingly important as the axion density becomes large. It is thus plausible to ask whether these higher-order terms, some of which give rise to repulsive self-interactions, can stabilize the collapsing axion star prior to its formation of a black hole state. In this note, we will consider the consequences of including the full self-interacting axion potential in the collapse of heavy, weakly bound axion stars.

In Section \ref{SecExpansion}, we review the nonrelativistic limit of axion field theory in the description of axion stars; then in Section \ref{SecVariation}, we outline the variational method used to find energetically stable bound states, and the computation of the total collapse time for large mass solutions. We estimate the binding energy in Section \ref{SecBinding}, in the initial and final states, but also dynamically in time during collapse. As the binding energy increases, it is known \cite{Lifetime} that the rate of decay for axion stars, through an annihilation process which ejects relativistic axions, rises quickly. We thus investigate whether collapsing axion stars emit a large fraction of their energy and decay due to quantum mechanical effects. Finally, we outline our conclusions in Section \ref{SecConclusions}.

\section{The Non-Relativistic Expansion for Axion Stars} \label{SecExpansion}
Axions are real scalar fields, but in the nonrelativistic limit can be described by a complex wavefunction $\psi$, using the expansion \cite{Guth,Nambu}
\begin{equation} \label{expansion}
 \phi = \frac{1}{\sqrt{2m}}\Big[e^{-i\,m\,t}\psi + e^{i\,m\,t}\psi^*\Big],
\end{equation}
which preserves the Hermiticity of the axion field $\phi$. At low temperatures, the wavefunction $\psi$ describes collectively a condensed state of $N$ axions, termed an axion star, and is normalized as $\int d^3 r |\psi^*\psi| = N$. The Klein-Gordon equation for $\phi$, expanded using eq. (\ref{expansion}) in the non-relativistic limit, yields the Lagrangian density
\begin{align}
 \Lagr &= \frac{1}{2}\partial_\mu\phi\partial^\mu\phi - V(\phi) \nn \\
 	&= i\psi^*\dot{\psi} - i \dot{\psi}^*\psi - 
	  \Big[\frac{|\nabla\psi|^2}{2m} + W(\phi)\Big]
\end{align}
for $\psi$, where 
\begin{equation} \label{InstPot}
 V(\phi) = m^2\,f^2\Big[1-\cos\Big(\frac{\phi}{f}\Big)\Big] 
\end{equation}
is the low-energy axion potential, with $m$ and $f$ the mass and decay constant of the axion, respectively. The gravitational potential 
\[
 V_{grav}(|\psi|^2) = -G\,m^2\int 
	\frac{\psi^*(x^\prime)\psi(x^\prime)}{|x^\prime-x|}d^3x^\prime,
\]
 representing the self-gravity of the condensate, can be added by hand \cite{Guth,Toth}. Then the quantity
\begin{equation} \label{Hamiltonian}
 H = \int d^3r\Big[\sum_i p_i\,\dot{q}_i - \Lagr\Big] 
	  = \int d^3r\,\Big[\frac{|\nabla\psi|^2}{2m} + W(\phi) 
		  + \frac{1}{2}V_{grav}(|\psi|^2)\Big]
\end{equation}
is conserved. Here $W(\phi)$ describes the quantum self-interactions of the axion field,
\begin{align} \label{potential}
 W(\phi) &= m^2\,f^2\,\Big[1 - \cos\Big(\frac{\phi}{f}\Big)\Big] 
		-\frac{m^2}{2}\phi^2 \nn \\
	 &= - m^2\,f^2\,\sum_{n=2}^{\infty}\frac{(-1)^n}{(2n)!}\Big(\frac{\phi}{f}\Big)^{2n}.
\end{align}
Note that the mass term in the first equality of eq. (\ref{potential}) is included to account for a cancellation in the non-relativistic limit between the potential and kinetic terms in $\Lagr$. In the nonrelativistic limit, the total energy per axion is $E_{tot}/N \simeq m$; that is, the binding energy $E_{tot}/N - m$ in the axion star is small. In that case, we can expand eq. (\ref{potential}) using eq. (\ref{expansion}) and drop the rapidly oscillating terms containing extra factors of $e^{\pm i m t}$. The resulting equation of motion for $\psi$ is the nonlinear Schr\"odinger equation.

We derive the total energy from eq. (\ref{Hamiltonian}) in the following way. The $n$th term in eq. (\ref{potential}) contains the factor
\[
 \phi^{2n} = \frac{^{2n}C_n}{(2m)^n}(\psi^*\,\psi)^n + \order(e^{\pm i\,m\,t}),
\]
where $^{2n}C_n$ are binomial coefficients. Dropping the rapidly oscillating pieces, we obtain
\begin{align} \label{SIpotential}
 W(\phi) &= - m^2\,f^2\,\sum_{n=2}^{\infty}
	  \frac{^{2n}C_n\,(-1)^n}{(2n)!}
	      \Big(\frac{\psi^*\,\psi}{2 m f^2}\Big)^{n} \nn \\
  &= m^2\,f^2\Big[1 - \frac{\psi^*\,\psi}{2 m f^2} 
	  - \sum_{n=0}^{\infty} \frac{(-1)^n}{(n!)^2}
	      \Big(\frac{\psi^*\,\psi}{2 m f^2}\Big)^{n} \Big] \nn \\
  &= m^2\,f^2\Big[1 - \frac{\psi^*\,\psi}{2 m f^2}
	  - J_0\Big(\sqrt{\frac{2\psi^*\,\psi}{m f^2}}\Big) \Big]
\end{align}
where $J_0(x)$ is a Bessel function of the first kind.\footnote{The $J_0$ dependence of the axion self-interaction potential was pointed out in \cite{ESVW} and later in \cite{Braaten,Braaten2}} Including also the kinetic and gravitational terms, the total energy functional has the form
\begin{equation} \label{Energy}
 E(\psi) = \int d^3r \Big[\frac{1}{2m}|\nabla \psi|^2 
	+ \frac{1}{2} V_{grav}\,|\psi^*\psi| 
	+ m^2\,f^2\,\Big(1 - J_0\big(\sqrt{\frac{2\psi^*\psi}{m\,f^2}}\big)\Big)
	- \frac{m}{2}\psi^*\psi \Big].
\end{equation}

A minimum of the energy correponds to a stable bound state, an axion star. Typically, one expands the Bessel function in eq. (\ref{Energy}) to obtain the leading self-interaction term, which is proportional to $(\psi^*\psi)^2$. This leading self-interaction is $\emph{attractive}$, and as we will explain below, this implies that the potential appears unbounded from below as the axion star size decreases, $R\rightarrow 0$. There can nonetheless exist local energy minima, corresponding to metastable states which are dilute and weakly bound. However, there exists a critical particle number $N_c$ above which no energy minimum exists, local or global. As a result, it is often assumed that an axion star with $M > m\,N_c$, being gravitationally unstable, will collapse all the way to a black hole state. A full description of this process can be found in \cite{ChavanisCollapse}, who used a Gaussian ansatz for the wavefunction and calculated the time for collapse to a black hole, which was on the order of an hour.

The full axion self-interaction potential, given by eq. (\ref{SIpotential}), contains additional terms beyond the attractive $(\psi^*\psi)^2$, which depend on increasing powers of the field $\psi$. Indeed, these higher-order terms, beginning with a repulsive $(\psi^*\psi)^3$ term, become increasingly relevant as the system increases in density, and we wish to investigate whether these terms have the effect of stabilizing the potential against complete collapse. To this end, we will examine the energy functional, including higher-order interactions, and determine whether the endpoint of collapse can lie at a radius greater than the Schwarzschild radius of the axion star. Such a result would be evidence that axion stars stabilize before they collapse to black holes.

\section{Variational Method} \label{SecVariation}
We will use a variational ansatz for the wavefunction to calculate the energy in eq. (\ref{Energy}) as a function of the condensate size, in order to estimate the positions of any energy minima. Using the result of \cite{ESVW}, we know how the macroscopic parameters of a weakly bound axion star, the radius $R$ and the axion number $N$, scale with the dimensionful parameters of the theory; we thus define the dimensionless quantities $\rho$ and $n$ by
\begin{equation} \label{scaling}
 R = \frac{1}{m}\,\frac{\rho }{\sqrt{\delta}} \qquad N = \frac{f^2}{m^2}\,\frac{n }{\sqrt{\delta}},
\end{equation}
where $\delta \equiv f^2/M_P^2$ and $M_P=G^{-1/2}$ is the Planck mass. For QCD axions, typical values are $m=10^{-5}$ eV and $f = 6\times10^{11}$ GeV, implying 
$\delta=\mathcal{O}(10^{-14})$ \cite{Cortona}.
 
 We will use a single variational parameter, the rescaled radius $\rho$, at fixed rescaled axion number $n$. Then the general form of a variational ansatz will be
\begin{equation}\label{ansatz}
\psi(r)= w\, F\left(\frac{r}{R}\right)\equiv w\,F(\xi),
\end{equation}
where at fixed $\xi$ the function $F(\xi)$ is independent of $\rho$, $n$, and $\delta$.  Then substituting the ansatz into the normalization condition of $\psi$ gives the normalization constant as
\begin{equation}
w=\sqrt{\frac{\delta\,n\,m}{C_2}} \frac{f}{\rho^{3/2}},
\end{equation}
where we introduced the notation
\begin{equation}
C_k=4\,\pi\int d\xi\,\xi^2 \,F(\xi)^k.
\end{equation}

Using (\ref{Energy})  we obtain for the energy functional
\begin{equation}\label{binding}
 \frac{E(\rho)}{m\,N} = \delta\left(\frac{D_2}{2\,C_2}\frac{1}{\rho^2}-\frac{B_4}{2\,C_2{}^2}\,\frac{n}{\rho}-\frac{n}{\rho^3}\,v\right),
 \end{equation}
where 
\begin{align}\label{vform}
v&=4\,\pi\,\frac{\rho^6}{n^2\,\delta^2}\int   d\xi\,\xi^2\left[1-J_0\left(\sqrt{\frac{2\,n\,\delta}{C_2\,\rho^3}}F(\xi)\right)-\frac{n\,\delta}{2\, C_2\,\rho^3}F(\xi)^2\right]\\
&=\sum_{k=0}^\infty \left(-\frac{1}{2\,C_2}\right)^{k+2}\left(\frac{n\,\delta}{\rho^3}\right)^k\frac{C_{2\,k+4}}{[(k+2)!]^2},
\end{align}
and where we defined the functions
\begin{align}
D_2&=4\,\pi\int d\xi\,\xi^2 \,F'(\xi)^2,\\
B_4&=32\,\pi^2\int d\xi\,\xi \,F(\xi)^2\int_0^\xi d\eta\, \eta^2\,F(\eta)^2.
\end{align}
Note that just like $C_k$, also $B_4$ and $D_2$ are independent of the physical parameters $n$, $\rho$, and $\delta$. The leading-order approximation of (\ref{binding}) is obtained when we take the small $\delta$ limit, at which
\begin{equation}\label{deltaeq0}
v_0=v|_{\delta=0}=\frac{C_4}{16\,C_2{}^2}.
\end{equation}

The minimization of (\ref{binding}) with respect to $\rho$ locates the radii of metastable minima and maxima of the binding energy.  The condition for the existence of metastable states constrains the reduced particle number $n$ to a finite constant.  Restricting ourselves to leading-order of $\delta$, which is $\delta=\mathcal{O}(10^{-14})$ in QCD (for $f = 6\times10^{11}$ GeV and $m=10^{-5}$ eV),  we obtain 
\begin{equation}
n_{c}=\sqrt{\frac{8}{3}}\frac{C_2\,D_2}{\sqrt{B_4\,C_4}}.
\end{equation}
For $n< n_c$, there exists a metastable minimum of the energy at a reduced radius
\begin{equation}
 \rho_{min} = \frac{C_2\,D_2}{B_4\,n}\Big[1 - \sqrt{1 - \frac{3}{8}\frac{B_4\,C_4}{C_2{}^2\,D_2{}^2} n^2}\Big]
\end{equation}
which, at $n=n_c$, has a value of
\begin{equation}
 \rho_* \equiv \rho_{min}\Big|_{n=n_c} = \sqrt{\frac{3\,C_4}{8\,B_4}}.
\end{equation}

 \subsection{Gaussian Ansatz}
 Following \cite{ChavanisMR,ChavanisCollapse}, we use a Gaussian ansatz to approximate the axion star wavefunction:
 \begin{equation} \label{GausAnsatz}
  \psi(r) = \frac{\sqrt{N}}{\pi^{3/4}\sigma^{3/2}} e^{-r^2/2\sigma^2},
 \end{equation}
 which corresponds to eq. (\ref{ansatz}) with $w = \sqrt{N}/(\pi^{3/4}\sigma^{3/2})$ and $F(\xi) = e^{-\xi^2/2}$.
 Note that when we talk about the ``size'' of such a condensate (whose wavefunction extends to $r\rightarrow\infty$), we refer to the conventional $R_{99}$, inside which $.99$ of the mass is contained. For the Gaussian ansatz, this occurs not at $\sigma$, but at a value closer to $3\sigma$. Note also that we define $\rho$ below using eq. (\ref{scaling}) with $R=\sigma$, not $R=R_{99}$.
 
\begin{figure}[ht]
   \includegraphics[scale=.75]{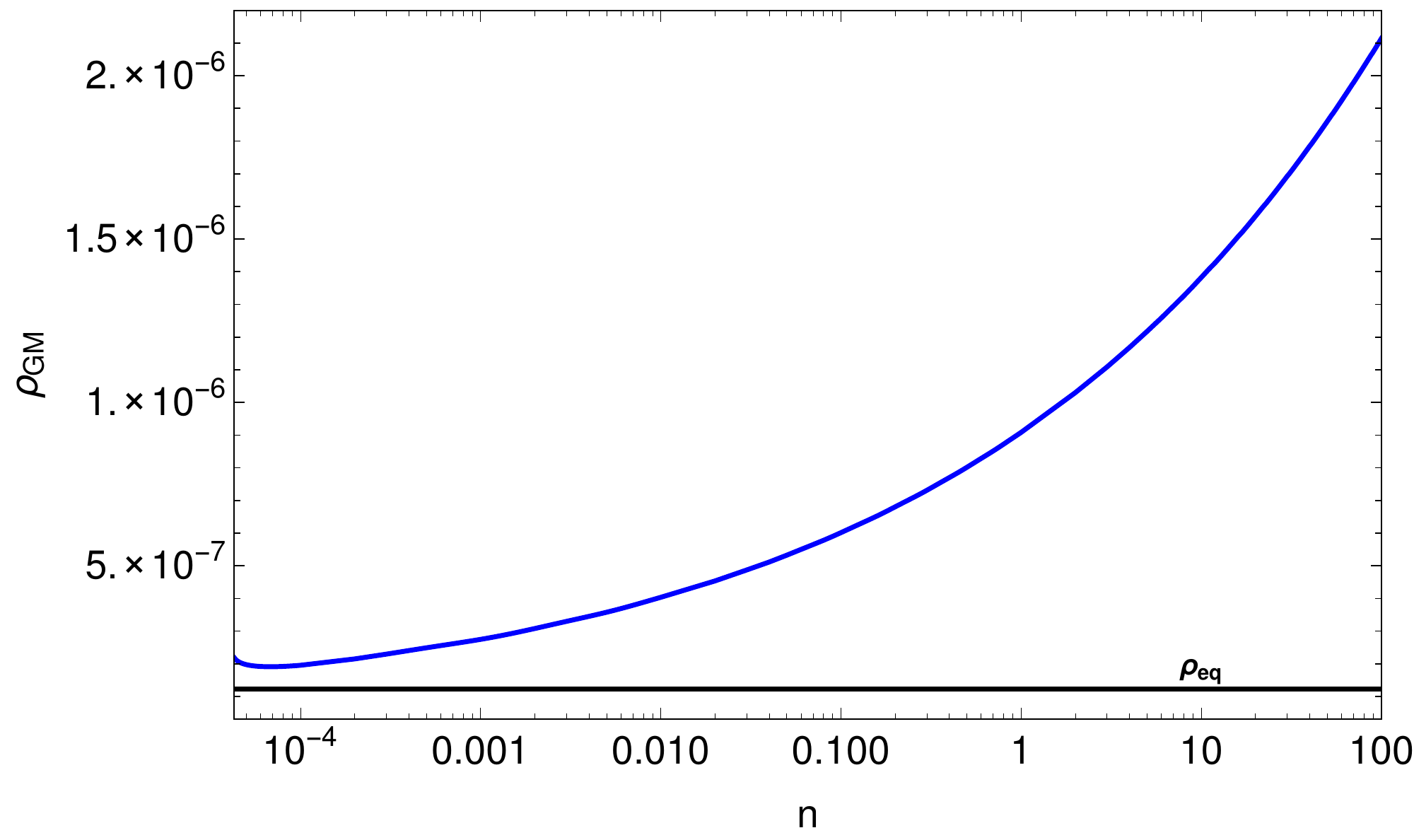}
   \caption{The position $\rho_{GM}$ of the global minimum of the energy in eq. (\ref{EnergyGausInt}), as a function of the reduced particle number $n$. The global minimum always lies at a radius $\rho_{GM}>\rho_{eq}$, the position at which the kinetic energy $\sim 1/\rho^2$ becomes dominant, which is depicted by the horizontal black line.}
   \label{rhoGMfig}
  \end{figure}

 The energy functional, given by eq. (\ref{binding}), depends on the coefficients
 \begin{align}
  D_2 = \frac{3\pi^{3/2}}{2}, \qquad B_4 = \sqrt{2\pi^5}, \qquad C_k = 2\sqrt{\frac{2\pi^3}{k^3}},
 \end{align}
 computed using the Gaussian function in eq. (\ref{GausAnsatz}). Written out explicitly, we have
 \begin{align}
  \frac{E(\rho)}{m\,N} &= \frac{3}{4}\frac{\delta}{\rho^2}
	- \frac{1}{\sqrt{2\pi}}\frac{n\,\delta}{\rho}\
	+ \int_0^\infty \frac{4\pi x^2}{n\,\delta}\Big[
	    1-J_0\Big(\sqrt{\frac{2n\,\delta}{\pi^{3/2}\rho^3}}
	    e^{-x^2/2\rho^2}\Big)
	- \frac{n\,\delta}{2\pi^{3/2}\,\rho^3}e^{-x^2/\rho^2}\Big]dx  
	      \label{EnergyGausInt} \\
      &= \delta\Big[\frac{3}{4}\frac{1}{\rho^2}
	- \frac{1}{\sqrt{2\pi}}\frac{n}{\rho}
	- \frac{1}{2\,\delta}\sum_{k=0}^{\infty}\frac{(-1)^k}{[(k+2)!]^2\,(k+2)^{3/2}}\Big(\frac{n\,\delta}{2\pi^{3/2}\rho^3}\Big)^{k+1}\Big],
	   \label{EnergyGaus}
 \end{align}
 where in the second equality we have expanded $J_0$ and integrated term by term. Though no closed form exists for the integral in eq. (\ref{EnergyGausInt}), we show in the Appendix that it is finite as $\rho\rightarrow0$, and that the kinetic energy term (which is proportional to $1/\rho^2$) dominates in this region. Consequently, the total energy is bounded from below in this formalism, and always has a global minimum.\footnote{It is possible that this conclusion would be modified by post-Newtonian corrections to the gravitational interaction.} In Figure \ref{rhoGMfig}, we show the position $\rho_{GM}$ of this global minimum of the energy functional in eq. (\ref{EnergyGausInt}) as a function of $n$. The global minimum always lies at a very small radius: $\rho_{GM}=\mathcal{O}(10^{-7}-10^{-6})$ depicted in the plot correspond to $R_{GM} \sim 5-50$ cm. Given that the Schwarzschild radius $R_S = 2\,M/M_P^2$, the ratio $R/R_S = \rho/(2\,n\,\delta)$. Thus
  \[
   \rho_S = 2\,n\,\delta
  \]
  so at $n=n_c$, $\rho_S = 10^{-13} \ll 10^{-7} < \rho_{GM}$, and this possible endpoint of collapse is not a black hole.
 
 The normalized energy per particle coming from the self-interaction is shown to be a constant $-1/2$ in the small $\rho$ limit, so we can estimate the value of $\rho\ll1$ at which the kinetic and self-interaction energies are of the same order; we find comparable magnitudes
 \[
  \frac{3}{4}\frac{\delta}{\rho^2} \sim \frac{1}{2}
 \]
 at a radius of $\rho_{eq} \sim 10^{-7}$, corresonding to roughly $R_{eq}\sim 5$ cm. $\rho_{eq}$ is shown as a horizontal black line in Figure \ref{rhoGMfig}.   This radius $R_{eq}$ is of the same order as the axion reduced Compton wavelength, $\lambda_c = \hbar/m\,c \sim 2$ cm. It should be noted that on length scales of $\mathcal{O}(\lambda_c)$, neglecting higher powers of $e^{\pm imt}$ in the expansion of eq. (\ref{expansion}) would fail, as special relativistic corrections to the kinetic energy could be large. Nonetheless, weakly bound stars have radii much larger than this, and as we describe below, even collapsing stars are well described by the non-relativistic approximation until the last moments of collapse. We have estimated the leading correction to the kinetic energy, which is $\sim p^4$, and in the range of $\rho$ considered here, its expectation value is down by a factor proportional to $\delta/\rho^2\ll 1$ compared to the leading-order term. We will thus postpone any further consideration of these relativistic corrections to the energy, which will be addressed in a future publication.
 
 In this work we analyze the low-energy axion potential in eq. (\ref{InstPot}), sometimes called the instanton potential. But it is well-known (see e.g. \cite{Cortona,Vecchia}) that an improved approximation is the chiral potential
 \[
  V(\phi) = m_\pi^2\,f_\pi^2\Big[1 - \sqrt{1 
      - \frac{4\,m_u\,m_d}{(m_u+m_d)^2}\sin^2\Big(\frac{\phi}{2\,f}\Big)}\Big],
 \]
 where $m_\pi$ and $f_\pi$ are the mass and decay constant of the QCD pion. This expression takes into account the non-perturbative effects of up and down quark masses $m_u$ and $m_d$. We find that substituting eq. (\ref{InstPot}) with this chiral potential does not qualitatively change the conclusions of this work: the global minimum of the energy in Figure \ref{rhoGMfig} shifts down by at most a few percent, still significantly larger than the corresponding black hole state. We put off any further discussion of the chiral potential to a future publication.

 We also consider the effect of including a finite but increasing number of terms in the series of eq. (\ref{EnergyGaus}). Because it has no closed form resummation, what is typically done is to truncate the series at some maximum $k=K$. We denote the truncated energy by $E_K(\rho)$, so that $\lim_{K\rightarrow\infty}E_K(\rho) = E(\rho)$. We also define a dimensionless truncated energy
 \begin{equation} \label{eKeq}
  e_K(\rho) \equiv \frac{E_K(\rho)}{m\,N\,\delta} 
	= \frac{3}{4}\frac{1}{\rho^2}
	- \frac{1}{\sqrt{2\pi}}\frac{n}{\rho}
	- \frac{1}{2\,\delta}\sum_{k=0}^{K}\frac{(-1)^k}{[(k+2)!]^2\,(k+2)^{3/2}}\Big(\frac{n\,\delta}{2\pi^{3/2}\rho^3}\Big)^{k+1}\Big].
 \end{equation}
 The minima of $e_K(\rho)$ should, at sufficiently large $K$, approximate well the stable bound states of the full energy function.
 
  The existence of a global minimum of the full energy functional in eq. (\ref{EnergyGaus}) has important consequences. In particular, we have pointed out above that this minimum lies at a radius many orders of magnitude larger than the Schwarzschild radius of the axion star, providing evidence that such objects do not collapse to black holes. Further, we note that the terms contained in the series of eq. (\ref{eKeq}) alternate between attractive and repulsive interactions, even and odd $k$ respectively. But as a result, a truncated energy $e_K(\rho)$ in eq. (\ref{eKeq}) for any even $K$ has no global minimum, and thus such a truncation removes the possiblity of approximating the stable radius of the full energy functional. We thus submit that when considering dense configurations of axions or collapse of axion stars, it is important to truncate the series on a repulsive term to preserve the global minimum.   
  
       \begin{figure}[ht]
   \includegraphics[scale=1]{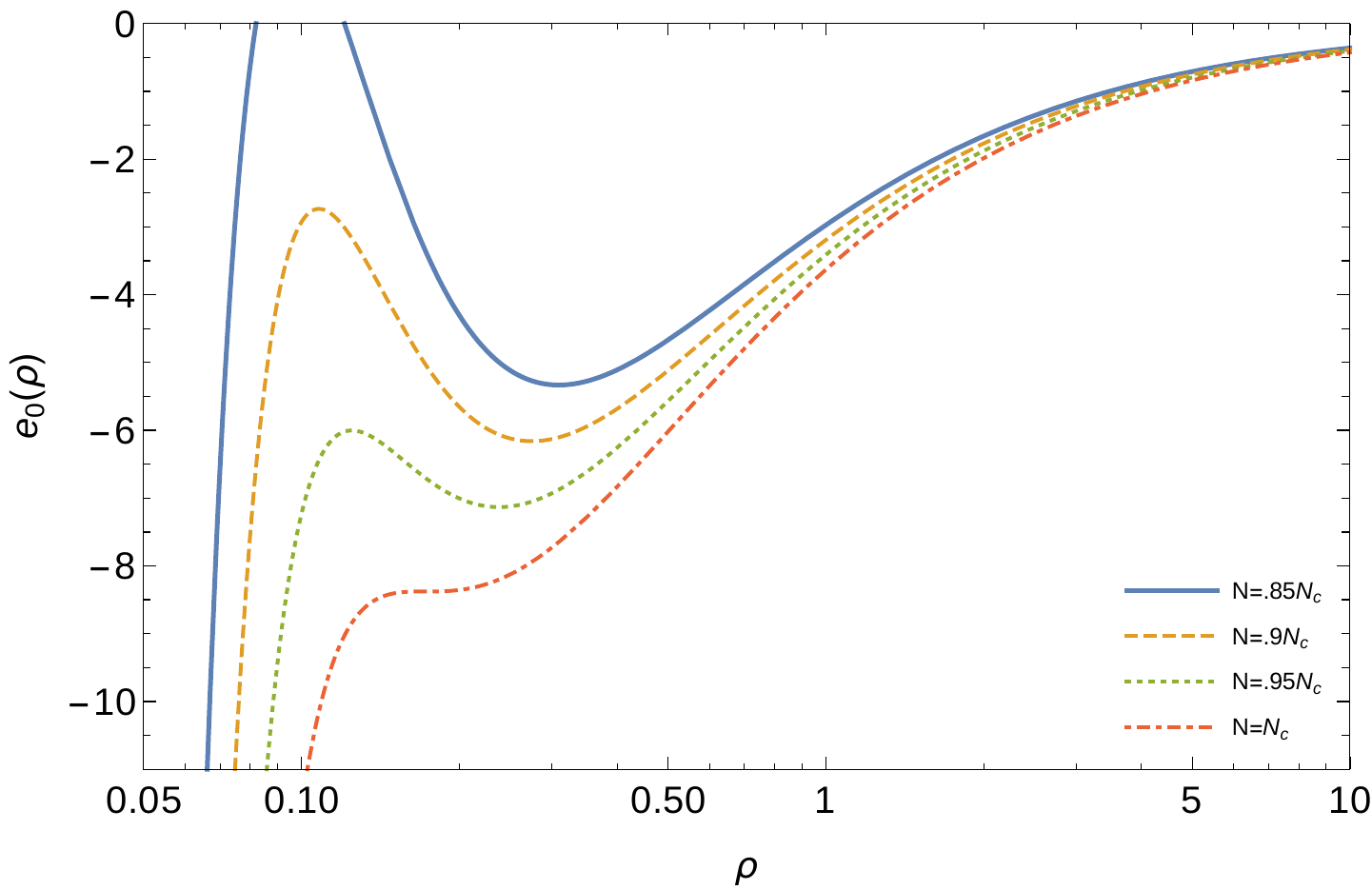}
   \caption{The truncated energy $e_0(\rho)$ near the position of the dilute minimum $\rho_*$ for different choices of particle number: $N=.85N_c$, $N=.9N_c$, $N=.95N_c$, and $N=N_c$. Note that the local minimum at $\rho_*$, represented in the plot, disappears at $N=N_c$. Including additional terms in $e_K(\rho)$ for $K>0$ makes a negligible difference in this range of $\rho$.}
   \label{diluteFig}
  \end{figure}
 
  \begin{figure}[ht]
   \includegraphics[scale=1]{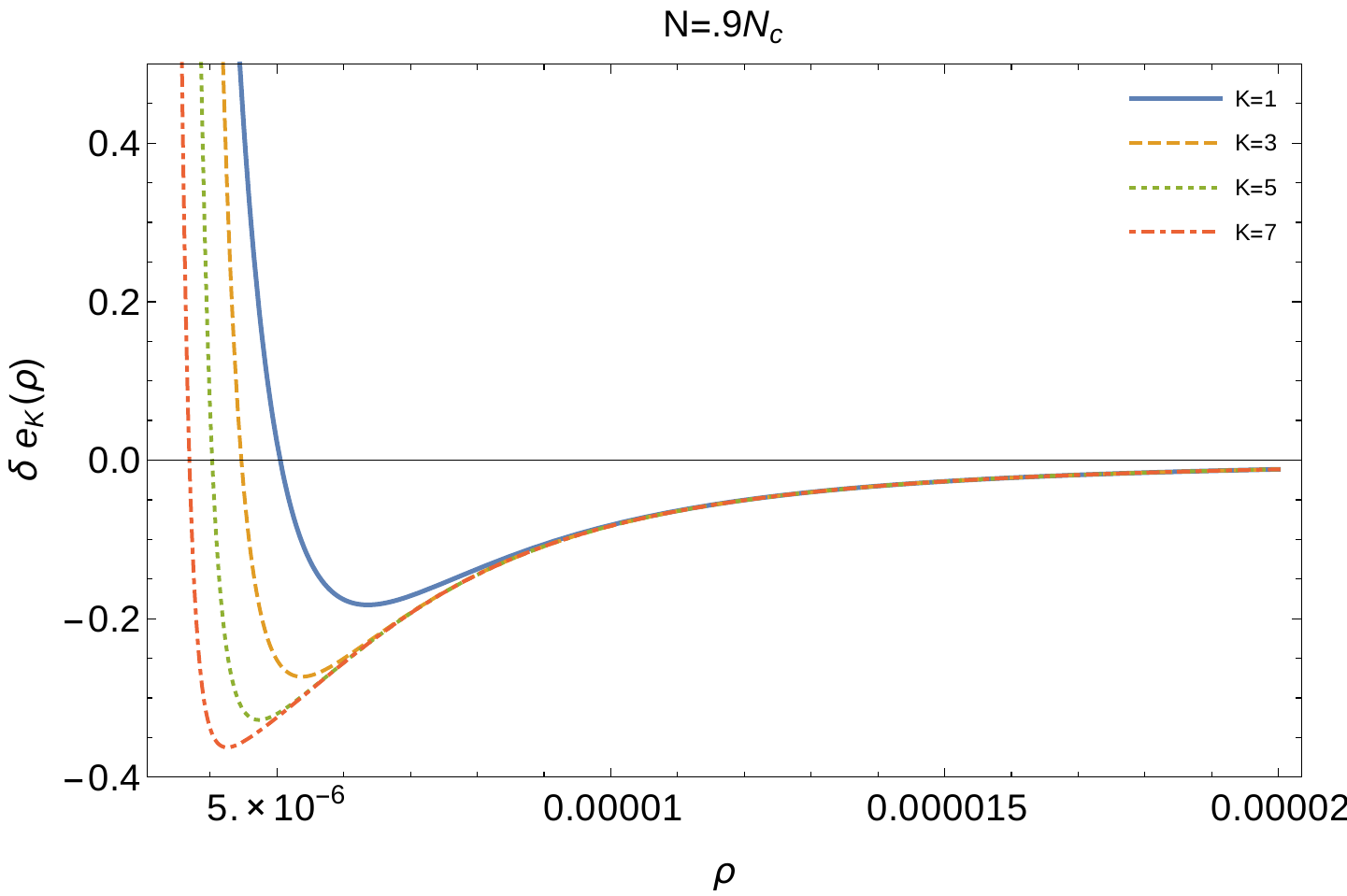}
   \caption{The energy $e_K(\rho)$ multiplied by the small parameter $\delta$ for $N=.9N_c$ at increasing odd orders in $K$: $K=1$, $K=3$, $K=5$, and $K=7$. The existence of a dense global energy minimum is preserved at each order, but shifts to smaller radii as $K$ increases. The repulsive kinetic term $\sim1/\rho^2$ dominates the total energy at $\rho=\rho_{eq}\sim10^{-7}$.}
   \label{orderFig}
 \end{figure}
 
 The leading-order interaction term is contained in $e_0(\rho)$, and has been considered in great detail previously \cite{ESVW,Guth,ChavanisMR,ChavanisCollapse}. It has been pointed out that there exists a maximum particle number $N=N_c$ above which no stable energy minimum exists. This critical value corresponds to a radius of $R_{99}\sim 500$ km for QCD axions \cite{ESVW}, and is approximated to the correct order of magnitude by the Gaussian ansatz, which gives a radius $R_{99}\sim 200$ km. In our notation, this critical particle number occurs at $n_c = 2\pi\sqrt{3}$ and at a radius $\rho_* = \sqrt{3/32\pi}$, for the Gaussian ansatz. We will use these as benchmark parameter values as we analyze the consequences of additional interaction terms in the axion potential. The energy functional in the vicinity of this minimum is shown in Figure \ref{diluteFig}.  It is also worth noting that the inclusion of additional terms in the self-interaction potential introduces negligible differences in this range of $\rho=\mathcal{O}(1)$; the leading expansion is an extremely good approximation in this region. But as noted above, any $e_K(\rho)$ for even $K$ (e.g. $e_0(\rho)$) is unbounded from below and will not be applicable in approximating the global energy minimum of the full potential, which is at $\rho\ll 1$.
 
 We turn now to $e_1(\rho)$, including the leading repulsive interaction which originates from a $(\psi^*\psi)^3$ term in the potential:
 \begin{equation} \label{e3eq}
  e_1(\rho) = \frac{3}{4}\frac{1}{\rho^2}
	- \frac{1}{\sqrt{2\pi}}\frac{n}{\rho} 
	    - \frac{1}{32\pi\sqrt{2\pi}}\frac{n}{\rho^3}
	    +\frac{\delta}{864\pi^3\sqrt{3}}\frac{n^2}{\rho^6}.
  \end{equation}
  In this case, the energy \emph{is} bounded from below and has a minimum at a very small radius $\rho = \rho_D$ (in contrast to the result using only $e_0$). At these small values of $\rho$, the energy is well approximated by the self-interaction terms only (gravity and kinetic energy are negligible); thus we can use the analytic expression
  \[
   \rho_D \approx \sqrt{\frac{2}{\pi}}\Big(\frac{n\,\delta}{3^{7/2}}\Big)^{1/3}
  \]
  to approximate the position of the global minimum. At $n=n_c$, $\rho_D\approx 7\times10^{-6}$, corresponding to $R_{99}\sim 7$ meters. Comparing with the global minimum of the full energy in Figure \ref{rhoGMfig}, we find a difference of only about a factor of $3-4$ near this value of $n\sim n_c$, a reasonable order of magnitude agreement. This justifies our truncation of the energy at the leading repulsive term, i.e. $e_1(\rho)$, in this analysis. The difference between $\rho_D$ and $\rho_{GM}$ does become large if $n$ increases far above $n_c$.
  
  We find that the existence of a dense global energy minimum is preserved at any odd $K$ in the approximation of eq. (\ref{eKeq}), and at increasing order, shifts to smaller radii (see Figure \ref{orderFig}). Nonetheless, the kinetic energy term dominates the full potential below $\rho_{eq}\sim10^{-7}$, and the global minimum of the full energy is at $\rho_{GM}>\rho_{eq}$, for any $n$. 
 
  The collapse of dark matter halos consisting of condensed scalar particles was examined by \cite{Harko}, using a time-dependent formalism that originated in \cite{PethikSmith}, and utilized by \cite{ChavanisMR,Perez}.
 The application of this method to an axion star, at leading-order in the self-interaction potential, was recently performed by \cite{ChavanisCollapse}. This collapse process is described by the dynamical equation
 \[
  E_{tot} = \alpha\frac{M}{2}\dot{R}(t)^2 - E(R)
 \]
 where $\alpha=3/4$ for the Gaussian ansatz, $E(R)$ is given by eq. (\ref{Energy}) and $E_{tot}$ is a constant. $R(t)$ is the size of the condensate, which varies with time during collapse. For a condensate with size $R_0$ at $t=0$, the time required to reach some other size $R(t)$ is given by
 \begin{align} \label{CollapseTime}
  t &= \sqrt{\frac{\alpha\,M}{2}} \int_{R(t)}^{R_0}\frac{dR}{\sqrt{E(R_0) - E(R)}} \nn \\
      &=\frac{M_P{}^2}{m\,f^2}\sqrt{\frac{\alpha}{2}}
      \int_{\rho(t)}^{\rho_0}\frac{d\rho}{\sqrt{e(\rho_0) - e(\rho)}},
 \end{align}
 where in the second equality we have rescaled the dimensionful quantities. 
 
 In the analysis of \cite{ChavanisCollapse}, $E(R)$ was approximated by the leading-order expression $E_0(R)$, and the collapse from $R_0=R_*$ to $R\rightarrow0$ was shown to last for a time which was on the order of an hour. We wish to investigate the effect of additional self-interactions in the axion potential on the collapse process. Including the first non-leading interaction piece, i.e. using $e_1(\rho)$, we have found that a global energy minimum exists at $\rho_D$; thus, we integrate eq. (\ref{CollapseTime}) not from $\rho=0$ but rather from $\rho=\rho_D$.
  
\begin{figure}[ht]
   \includegraphics[scale=1]{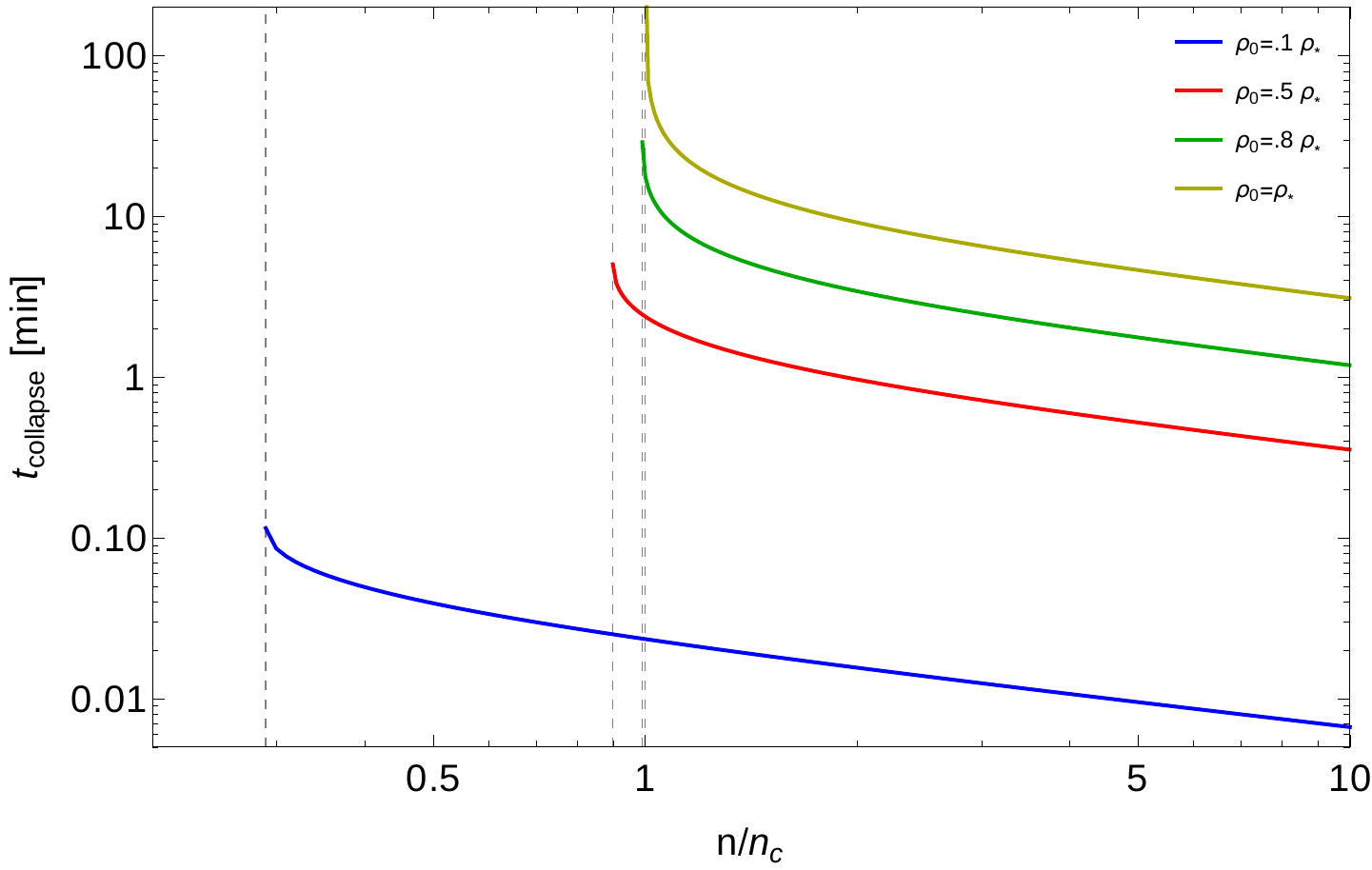}
   \caption{Collapse time for an axion star as a function of $n/n_c$, for different choices of starting radius $\rho_0$: $\rho_0 = .1\rho_*$, $\rho_0 = .5\rho_*$, $\rho_0 = .8\rho_*$, $\rho_0 = \rho_*$. At $N<N_c$, condensates can still collapse if the starting radius $\rho_0<\rho_*$.}
   \label{tCollapse}
  \end{figure}

  If the axion star begins its collapse at $\rho_0=\rho_*$, then of course at $n = n_c$ the collapse time is formally infinite, because the potential is flat at $\rho_*$. We consider values of $n$ which are slightly larger than $n_c$ and see how the collapse time changes. We also investigate the change in collapse time as the starting size $\rho_0$ deviates from $\rho_*$. This latter case could be of interest, say, if axion star collapse can be catalyzed by collisions with other astrophysical sources. In that case, even condensates with $N<N_c$ can collapse, provided some catalyzing interaction which reduces its initial radius to $R_0<R_*$. These considerations are represented together in Figure \ref{tCollapse}.
  
\begin{figure}[ht]
  \includegraphics[scale=1]{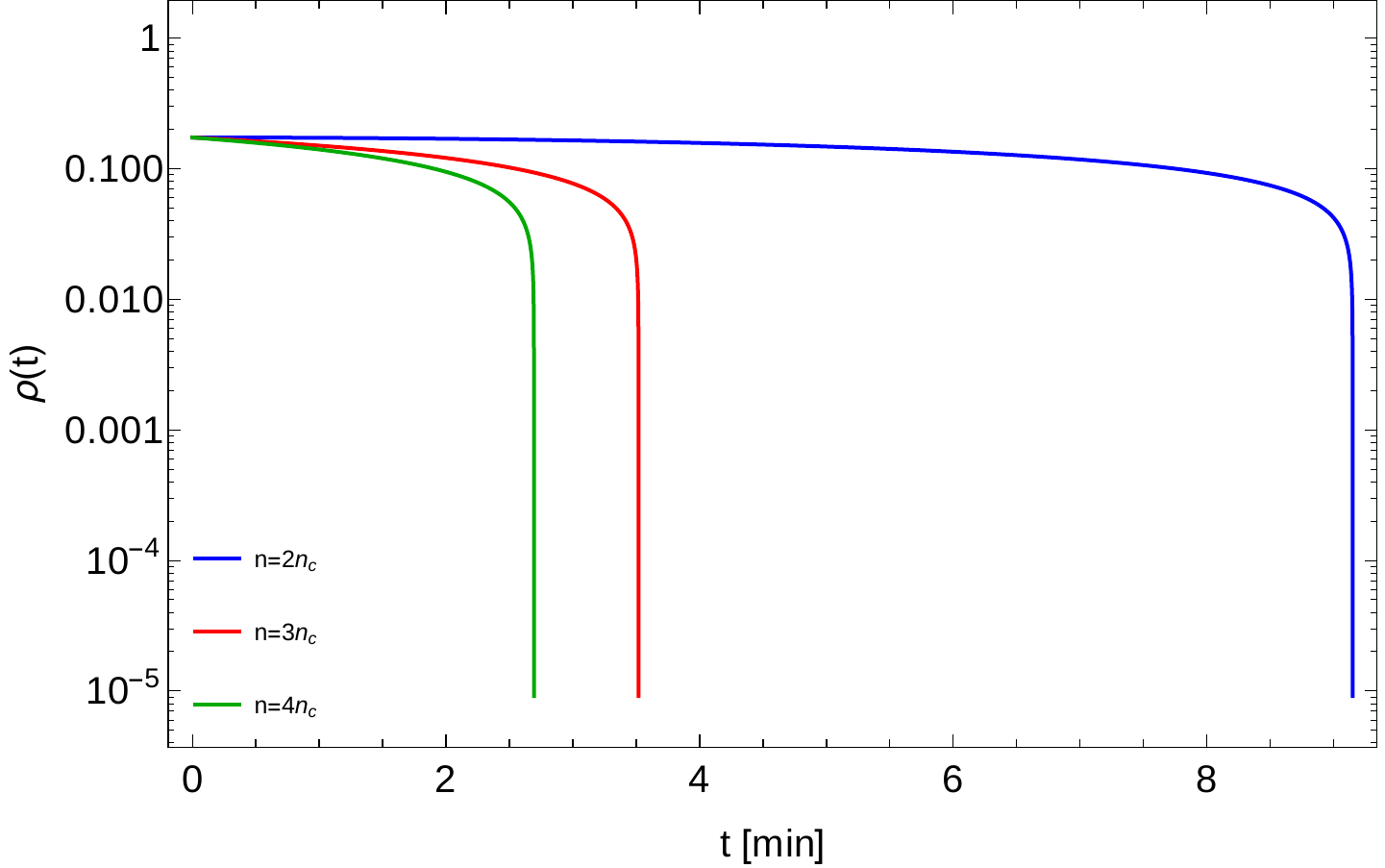}
  \caption{The dimensionless radius of a collapsing axion star using the approximate energy $E_1(\rho)$ as a function of time, for three choices of particle number $N$: $N=2N_c$, $N=3N_c$, and $N=4N_c$.}
  \label{rhooft}
  \end{figure}

   We can also track the radius of the axion star as a function of time, throughout the collapse process; see Figure \ref{rhooft}. For a large portion of the total collapse time, the radius changes little, as the star rolls slowly down a shallow potential, but later collapses fast to the dense minimum of radius $\rho_D$.
   
  \subsection{Cosine Ansatz}
\begin{figure}[ht]
   \includegraphics[scale=1]{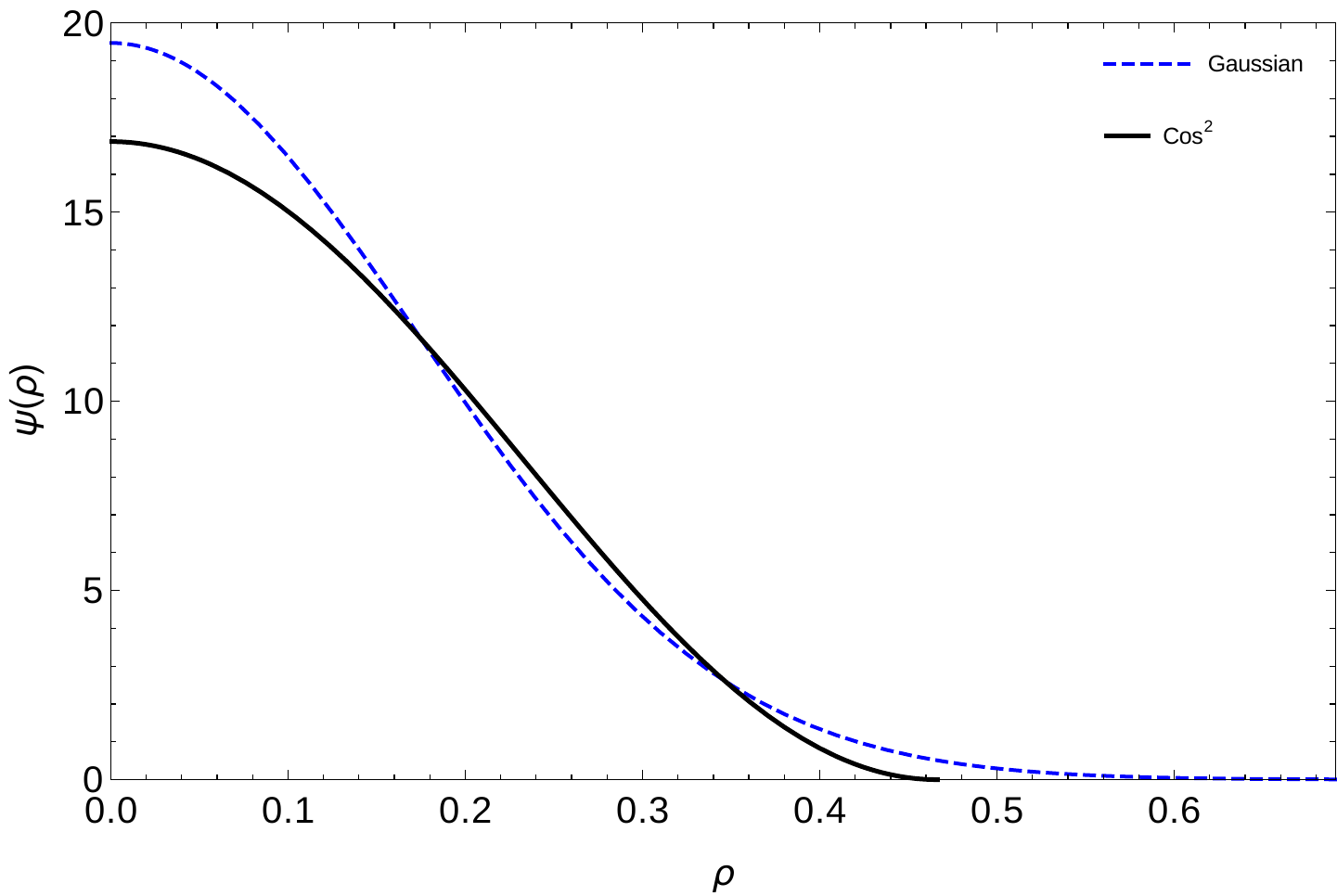}
   \caption{A comparison of the wavefunctions for the Gaussian ansatz (blue, dashed) and the cosine ansatz (black, solid), normalized to the same total size.}
   \label{ansatze}
 \end{figure}

  The Gaussian ansatz is believed to be a reasonable approximation to the axion star wavefunction. However, in order to verify that our results are not an artifact of the wavefunction one chooses, we present a second ansatz for the variational analysis:
  \[
   \psi(r) = \sqrt{\frac{4\,\pi\,N}{(2\pi^2-15)R^3}}\cos^2\Big(\frac{\pi\,r}{2\,R}\Big)
	\qquad (r < R).
  \]
  A comparison of the two ans\"atze we use is shown in Figure \ref{ansatze} for the same total size.\footnote{Note that while the $cos^2$ wavefunction goes to $0$ at some $r$ and thus has a definite edge, the Gaussian function (as we pointed out previously) does not.} The energy functional, rescaled and truncated as above, depends on the coefficients
  \begin{align}
   &D_2 =  \frac{\pi(2\pi^2-3)}{12},
	  \qquad B_4 = 8\pi\Big(\frac{3\pi}{80} - \frac{115}{768\pi}
		      - \frac{33341}{18432\pi^3}\Big), \nn \\
   C_2 = \frac{\pi}{2} - &\frac{15}{4\pi},
	  \qquad C_4 = \frac{35(24\pi^2-205)}{2304\pi},
	  \qquad C_6 = \frac{77(600\pi^2 - 5369)}{153600\pi}.
  \end{align}
  This implies that, for the cosine ansatz,
  \begin{align} \label{Ecosine}
   e_1(\rho) = &\frac{\pi^2(2\pi^2-3)}{6(2\pi^2-15)}\frac{1}{\rho^2}
	-\frac{(3456\pi^4-13800\pi^2-166705)}{1440(2\pi^2-15)^2}\frac{n}{\rho}
	    \nn \\
	&-\frac{35\pi(24\pi^2-205)}{2304(2\pi^2-15)^2}\frac{n}{\rho^3}
	+\frac{77\pi^2(600\pi^2-5369)}{691200(2\pi^2-15)^3}\frac{\delta\,n^2}{\rho^6}.
  \end{align}

  As before, we minimize the approximated energy $e_1(\rho)$ with respect to $\rho$, and find both a dilute and a dense minimum. The dilute minimum disappears above a critical particle number, corresponding to $n_c \approx 12.6$, where the radius is $\rho_* \approx .44$ (around $200$ km). The dense minimum is at approximately $\rho_D\approx7.4\times10^{-6} n^{1/3}$, within a factor of $2$ of the result from the Gaussian case, $\rho_D \approx 4.5\times10^{-6} n^{1/3}$.
  
 \section{Decay of Collapsing Solutions} \label{SecBinding} 
 In a previous work \cite{Lifetime}, some of us found that axion stars can decay through repeated occurrences of the a process which ejects relativistic axions from the star. Such a process is not forbidden by any symmetry because axions, being Hermitian fields, do not have a conserved number, and because bound axions, along with the axion star itself, are not in momentum eigenstates. To describe this interaction, the spectrum of bound states describing the axion star was extended by a collection of scattering states, labeled by momentum $p$. The leading contribution to this process was an interaction of the form $\mathcal{A}_N \rightarrow \mathcal{A}_{N-3} + a_p$, where $\mathcal{A}_N$ denotes an axion star with $N$ axions and $a_p$ denotes a relativistic axion with momentum $p$. Without the addition of these scattering states, the matrix element for this and many other interactions are identically zero. Our analysis assumed a small binding energy in the axion star. A contrarian point of view was expressed in \cite{Braaten2016}.
 
 We found in \cite{Lifetime} that the lifetime of an axion star through emission of relativistic axions depends on a reduced binding energy parameter $\Delta \equiv \sqrt{1 - (E_{tot}/N\,m)^2}$. The leading-order expansion in $\Delta\ll 1$ is equivalent to the infrared limit of the theory, where only the marginal $\phi^4$ term appears in the interaction potential \cite{ESVW}. For weakly bound stars, the leading process $\mathcal{A}_N \rightarrow \mathcal{A}_{N-3}+a_p$ has a rate which, as a function of $\Delta$, is dominated by an exponential factor,
 \begin{equation} \label{decayrate}
  \Gamma = \frac{f^2}{2\sqrt{8} \pi m}\Big[\frac{32\pi r}{3\Delta}
	  \exp\Big(-\frac{\sqrt{8}r}{\Delta}\Big)\Big]^2
 \end{equation}
 with $r=.603156$. Axion stars with masses near the maximum have very small binding energies, corresponding to $\Delta=\mathcal{O}(10^{-7})$, and are thus very stable in this sense. More generally, we found that if a star has $\Delta \lesssim .05-.06$, then it is stable on timescales as long as the age of the universe, because the lifetime
 \begin{equation} \label{Lifetime}
  \tau = \frac{3 y_M}{1024\pi r^3}\frac{\Delta^2}{m} \exp\Big(\frac{2\sqrt{8}r}{\Delta}\Big),
 \end{equation}
 is a monotonically decreasing function of $\Delta$ in the relevant range. The constant in eq. (\ref{Lifetime}) has the value $y_M = 25.46$.
 
 The dense energy minimum $\rho_D$ has a large binding energy, corresponding (in the Gaussian case) to $\Delta = .56$. A na\"ive application of eq. (\ref{Lifetime}) at this large value of $\Delta$ gives $\tau = 10^{-9}$ sec; however it is not known whether this estimate is reliable, since the analysis of \cite{Lifetime} applies only in the weak binding limit. Further, eq. (\ref{Lifetime}) takes into account only the attractive $\phi^4$ interaction, but this is a valid approximation throughout most of the decay process. Nonetheless, if valid, such a short lifetime would imply that these dense states, as the endpoint of collapse, would decay very quickly. However, our calculational method is not applicable to strongly bound systems, so we cannot make a definite statement about it. We hope to investigate the decay of strongly bound states in greater detail in the future. Recent investigations of collapse using a classical collapse analysis have concluded that collapsing axion stars lose a significant fraction of their mass through emission of relativistic axions \cite{Tkachev2016,Marsh}.
 
\begin{figure}[ht]
  \includegraphics[scale=1]{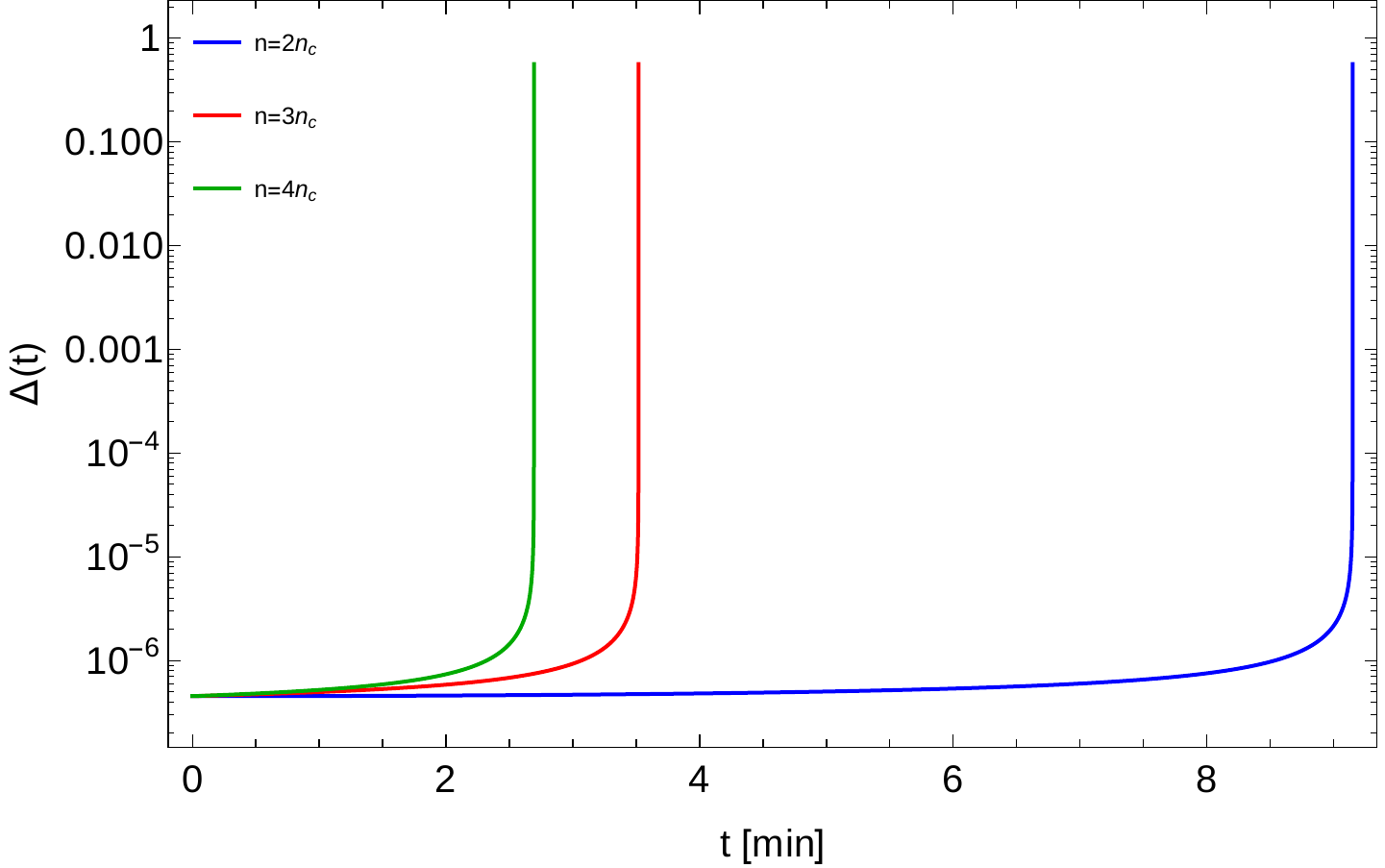}
    \caption{The reduced binding energy $\Delta$ of a collapsing axion star using the approximate energy $E_1(\rho)$ as a function of time, for three choices of particle number $N$: $N=2N_c$, $N=3N_c$, and $N=4N_c$.}
  \label{Deltaoft}
 \end{figure}

 In the weak binding region, where eq. (\ref{Lifetime}) holds, we know that $\Delta$ is a one-to-one function of $\rho$, and thus also of the collapse time $t$ as defined in eq. (\ref{CollapseTime}). We find that the binding energy obtains $\Delta \sim .05$ at $\rho\sim 10^{-4}$ (compared with $\rho_D\sim10^{-5}$). As a function of time, the binding energy only changes appreciably in the last fraction of a second of the collapse, but rises quickly to a strongly bound final state (see Figure \ref{Deltaoft}). In these last moments, the decay rate in eq. (\ref{decayrate}) becomes astronomically large; $\Gamma\sim 1$ emitted axion/sec at $\Delta\sim.0223$, and rises to $\Gamma\sim10^{50}$ emitted axions/sec at $\Delta\sim.1$. We therefore are led to the conclusion that axion stars, as they collapse, emit many highly energetic free axions.\footnote{If dark matter consists of axion stars, then this decay process could deplete the total amount of dark matter in galaxy clusters. This effect is considered in a different context in \cite{Tatsu}.} Such an explosion, referred to as a Bosenova, has been observed experimentally by condensed matter physicists using cold atoms \cite{Bosenova}. While this work was under review, a different group performing a numerical simulation also suggested that a large fraction of axion star energy is expelled during the collapse process through relativistic axion emission \cite{Tkachev2016}.
 
 We emphasize again that the analysis of the decay process in \cite{Lifetime} applies only at weak binding, when $\Delta\ll 1$. This condition holds for the dilute state as well as throughout a large portion of the collapse process, but it is possible that some new dynamics take hold at truly strong binding $\Delta=\mathcal{O}(1)$, as for the dense global minimum of the energy where $\Delta\sim.56$. We are led to the conclusion that relativistic axion emission becomes important during collapse, but it is possible that a stable, strongly bound remnant remains.
 
 \section{Conclusions} \label{SecConclusions}
 The contribution of the axion self-interaction potential to the total energy in the variational method can be computed to arbitrary order using an expansion in powers of the axion field. This expansion is equivalent to an expansion in the small parameter $\delta=f^2/M_P^2\ll1$. Because of the smallness of this parameter ($\delta \sim 10^{-14}$ for QCD axions), the potential is typically truncated at leading-order, including only the attractive $(\psi^*\psi)^2$ term. This truncation works extraordinarily well at large radii, and the dilute radius $R_*$ found by multiple authors previously \cite{ESVW,Guth} is preserved. In the regime of larger $\delta$ (e.g. axions with $f \sim .1\,M_P$), some of these conclusions could be changed. While this work was being reviewed, an analysis performed in the classical limit \cite{Marsh} suggested that axion theories with large $\delta$ indeed allow collapse to black holes in some regions of parameter space. We are working out the mass spectrum of axion stars in such theories, which will be the topic of future work.
 
 Going beyond the leading-order approximation, without truncation we have found that a global minimum of the full energy exists, which is not present in the leading-order expansion; we calculated its position, and it corresponds to a radius $R_{GM}$ many orders of magnitude larger than the corresponding Schwarzschild radius. We approximate this global minimum using a next-to-leading-order expansion, using the truncated energy of eq. (\ref{e3eq}), which has a global minimum at a radius $R_D\ll R_*$. For $m=10^{-5}$ eV QCD axions and using the Gaussian ansatz, the dilute radius $R_*\sim200$ km, while $R_D\sim 7$ meters. $R_D$ is a good order of magnitude estimate of $R_{GM}$.
 
 Previous analyses of collapsing boson stars with an attractive self-interaction have concluded (correctly) that, with nothing to stabilize the potential as $R\rightarrow 0$, the endpoint of collapse is a black hole state. For the axion potential, we have found higher-order self-interactions, some of which are repulsive, stabilize axion stars as they collapse and there exist energetically stable configurations at very small radii. These configurations correspond to dense axion star states which are nonetheless not black holes, and resemble closely the type of the dense states found by the authors of \cite{Braaten} using a different method. Dense configurations of this kind can exceed the maximum mass normally allowed for weakly bound axion stars, which is roughly $M_c\sim10^{19}$ kg for $m=10^{-5}$ eV axions \cite{ESVW}.
 
 We have examined the collapse dynamically in time, and find that masses $M$ just above $M_c$ collapse from $R=R_*$ in a time on the order of hours, or tens of minutes. The radius changes slowly at first, then drops rapidly as the slope of the potential becomes increasingly steep. Stars which begin collapse 
 at a radius $R_0<R_*$ were also considered, a case which is interesting if, for example, axion star collapse is catalyzed by collisions of two lighter axion stars. This could occur even if these stars do not become gravitationally bound to each other. This topic will be pursued in a future work.
 
 If stable, then heavy axion star states could be detectable via gravitational lensing experiments. Such states have large binding energies, and thus non-relativistic and non-perturbative corrections may become important in that regime. During collapse, however, when binding energies increase but are still sufficiently small, previous calculations \cite{Lifetime} suggest that the rate of emission of relativistic axions from an axion star will rise very rapidly. The rate of decay through the leading number-changing interaction $\mathcal{A}_N\rightarrow\mathcal{A}_{N-3}+a_p$ rises to $\Gamma \gtrsim 10^{50}$ emitted axions/sec in the final moments of collapse, leading to rapid emission of axions in what is often called a Bosenova \cite{Bosenova}. It is not clear in our analysis precisely what fraction of the energy of the star would be expelled through this process, or whether a stable dense state could remain. It would be interesting to investigate the energy spectrum of these collapses in detail, to determine if there are detectable consequences of such an explosion.

 \section*{Acknowledgements}
 We thank P. Argyres, R. Gass , A. Kagan, D. Kulkarni, J. Leeney, M. Ma, and C. Vaz for conversations. M.L. thanks the WISE program and Professor U. Ghia for support and encouragement, and the University of Cincinnati and the Department of Physics for a summer research fellowship. The work of JE was partially supported by a Mary J. Hanna Fellowship through the Department of Physics at University of Cincinnati, and also by the U.S. Department of Energy, Office of Science, Office of Workforce Development for Teachers and Scientists, Office of
Science Graduate Student Research (SCGSR) program. The SCGSR program is administered by the Oak Ridge Institute for Science and Education for the DOE under contract number DE-SC0014664.
 
 \section*{Appendix I: Total Energy as $\rho\rightarrow 0$}
 In this section we outline the proof that the contribution of the self-interaction potential to the total energy is finite in the $\rho\rightarrow0$ limit, and consequently that the kinetic energy $\sim 1/\rho^2$ is dominant. The self-interaction term, coming from eq. (\ref{EnergyGausInt}) for the Gaussian ansatz is of the form
 \begin{align}
  \frac{V_{SI}}{m\,N} &= \int_0^\infty\frac{4\pi x^2}{n\,\delta}\Big[1-J_0\Big(\sqrt{\frac{2n\,\delta}{\pi^{3/2}\rho^3}}
	    e^{-x^2/2\rho^2}\Big)
	- \frac{n\,\delta}{2\pi^{3/2}\,\rho^3}e^{-x^2/\rho^2}\Big]dx  \nn \\
	&= \frac{4\pi}{n\,\delta}\int_0^z\rho^3\sqrt{2\,\ln\frac{z}{u}}
		  \Big[1-J_0(u)-\frac{u^2}{4}\Big]\frac{du}{u} \nn \\
	&= \frac{4\pi}{n\,\delta}\,\mathcal{I} \label{intSI}
 \end{align}
 where in the second step we defined $u = z\,\exp(-x^2/2\rho^2)$ with $z=\sqrt{2n\delta/\pi^{3/2}\rho^3}$. We are interested in the case of $\rho\rightarrow0$, corresponding to $z\rightarrow\infty$.
 
 We proceed with the estimation of the integral $\mathcal{I}$ in the limit $\rho\rightarrow0$ in the following way. Break up the integral into two parts: (1) $I_1$, integrated over the interval $0<u<\nu$, where $1\ll\nu<z$, and (2) $I_2$, integrated over the remaining $\nu< u < z$. Consider first $I_1$: 
 at $u \ll 1/z$, the integrand is dominated by the expression
 \begin{align}
  \rho^3\sqrt{\ln\frac{1}{u}}\Big[1-J_0(u)-\frac{u^2}{4}\Big]\frac{1}{u}
	&\approx \rho^3\sqrt{\ln\frac{1}{u}}\Big[-\frac{u^4}{64}\Big]\frac{1}{u} \nn \\ 
	&= -\frac{\rho^3}{64}u^3\sqrt{\ln\frac{1}{u}} \nn
 \end{align}
 which goes quickly to $0$ as $u\rightarrow0$. At larger values $u\gg 1/z$, on the other hand, $I_1$ is dominated by the term
 \[
  \int_0^\nu \rho^3\sqrt{2\,\ln z}\Big[1-J_0(u)-\frac{u^2}{4}\Big]\frac{du}{u} 
	  \sim \rho^3 \sqrt{\ln z},
 \]
 and consequently,
 \begin{equation} \label{I1eq}
  I_1 \sim  \rho^3 \sqrt{\frac{1}{\rho}}
 \end{equation}

 For $I_2$, we consider large $u$, since $1\ll\nu < u$. Then the bracket in eq. (\ref{intSI}) dominated by the quadratic term $u^2/4$, since $J_0(u)\rightarrow0$. Thus,
 \[
 I_2 \approx -\sqrt{2}\rho^3\int_\nu^z \sqrt{\ln\frac{z}{u}}\frac{u}{4}du.
 \]
 Making the change of variables $t=\ln(z/u)$, we can write it in the form
 \begin{align}
  I_2 &\approx -\frac{\sqrt{2}}{4} \rho^3 z^2\int_0^{\ln(z/\nu)}\sqrt{t}e^{-2t} dt \nn \\
      &= -\frac{\sqrt{2}}{32}\rho^3 z^2 \Big[\sqrt{2\pi}\text{Erf}\Big(\sqrt{2\ln\frac{z}{\nu}}\Big)
		- 4\sqrt{\ln\frac{z}{\nu}}\big(\frac{\nu}{z}\big)^2\Big)\Big] \nn \\
      &= -\frac{n\,\delta}{8\pi}\Big[\text{Erf}\Big(\sqrt{2\ln\frac{z}{\nu}}\Big)
		- \frac{4\,\nu^2}{\sqrt{2\pi}}\frac{\sqrt{\ln\frac{z}{\nu}}}{z^2}\Big)\Big].
 \end{align}
 In the limit $z\rightarrow\infty$, the first term in the brackets $\rightarrow 1$, while the second term vanishes. Thus,
 \begin{equation} \label{I2eq}
  I_2 \approx -\frac{n\,\delta}{8\pi}
 \end{equation}
 
 Finally, since $I_1$, given by eq. (\ref{I1eq}), vanishes as $\rho\rightarrow0$, the self-interaction energy approaches a constant in the limit $\rho\rightarrow0$:
 \begin{equation}
  \frac{V_{SI}}{m\,N} \approx \frac{4\pi}{n\,\delta}\,I_2 
	  \approx -\frac{1}{2}.
 \end{equation}
 Because this result is finite, we are led to conclude that the kinetic energy, which diverges as $1/\rho^2$, provides the dominant contribution to the energy at $\rho\rightarrow0$. The gravitational interaction, which diverges as $-1/\rho$, is also negligible in this region. Thus the axion star energy is always bounded from below. Though we have here only proved this for the Gaussian ansatz, this conclusion is significantly more general, and will be investigated in detail in a future work.

\end{document}